\documentclass[a4paper,11pt]{article}
\pdfoutput=1

\usepackage{jheppub}
\usepackage{ctable}
\usepackage{float}
\usepackage{booktabs}
\usepackage{placeins}
\usepackage{multirow}

\def\tr{\mathop{\rm tr}\nolimits}

\def\Mod{\,{\rm mod}\,}

\newcommand\be{\begin{equation}}
\newcommand\ba{\begin{eqnarray}}
\newcommand\ee{\end{equation}}
\newcommand\ea{\end{eqnarray}}

\newcommand{\arxiv}[1]
  {\href{http://arxiv.org/abs/#1}{arXiv:#1}}
  
\newcommand{\arxivold}[1]
  {\href{http://arxiv.org/abs/#1}{#1}}

\definecolor{ourgray}{gray}{0.44}
\newcommand{\light}[1]{\textcolor{ourgray}{#1}}

\title{\boldmath Nonlinear Sigma Models with Compact Hyperbolic Target Spaces}

\author[a]{Steven Gubser}
\author[b,d]{Zain H.~Saleem}
\author[b]{Samuel S. Schoenholz}
\author[c]{Bogdan Stoica}
\author[b]{James Stokes}

\affiliation[a]{Joseph Henry Laboratories, Princeton University, \\
 Princeton, NJ  08544, USA}
\affiliation[b]{Department of Physics and Astronomy,\\University of Pennsylvania, Philadelphia, PA 19104}
\affiliation[c]{Walter Burke Institute for Theoretical Physics,\\ California Institute of Technology, 452-48, Pasadena, CA 91125, USA}
\affiliation[d]{National Center for Physics, \\ Quaid-e-Azam University Campus,
 Islamabad 4400, Pakistan}
\emailAdd{sgubser@princeton.edu}
\emailAdd{zains@sas.upenn.edu}
\emailAdd{schsam@sas.upenn.edu}
\emailAdd{bstoica@theory.caltech.edu}
\emailAdd{stokesj@sas.upenn.edu}

\preprint{CALT-TH 2015-019 \\ \hspace*{\fill} PUPT-2487}

\abstract{
We explore the phase structure of nonlinear sigma models with target spaces corresponding to compact quotients of hyperbolic space, focusing on the case of a hyperbolic genus-2 Riemann surface. The continuum theory of these models can be approximated by a lattice spin system which we simulate using Monte Carlo methods. The target space possesses interesting geometric and topological properties which are reflected in novel features of the sigma model. In particular, we observe a topological phase transition at a critical temperature, above which vortices proliferate, reminiscent of the Kosterlitz-Thouless phase transition in the $O(2)$ model \cite{KT1,KT2}.  Unlike in the $O(2)$ case, there are many different types of vortices, suggesting a possible analogy to the Hagedorn treatment of statistical mechanics of a proliferating number of hadron species.  Below the critical temperature the spins cluster around six special points in the target space known as Weierstrass points.  The diversity of compact hyperbolic manifolds suggests that our model is only the simplest example of a broad class of statistical mechanical models whose main features can be understood essentially in geometric terms.
}

\begin{document}
\maketitle
\flushbottom

\section{Introduction}
The breaking of continuous symmetries is accompanied by the appearance of massless Goldstone modes. Fluctuations of these modes destroy long-range order at any finite temperature in dimensions $d\leq 2$; this is the statement of the Mermin-Wagner theorem. This does not exclude the possibility of ``quasi long-range order'', however, in which correlators exhibit power-law, rather exponential decay. This behavior is associated with a continuous phase transition driven by the proliferation of topological defects~\cite{2dtransition}.

We can expect topological transitions to occur in a spin system whenever the homotopy group of the spin space is non-trivial. The XY model is the simplest example in which the spin space of $S^1$ has homotopy group $\pi_1(\mathbb{S}^1) = \mathbb{Z}$. This is to be compared with the $O(3)$-invariant Heisenberg model which has $\pi_1(\mathbb{S}^2) = 0$ and thus does not exhibit quasi-long-range order in $d\leq 2$. The first non-trivial example with two-dimensional spin space is the torus $\mathbb{T}^2 = \mathbb{S}^1 \times \mathbb{S}^1$ which is essentially two decoupled copies of the XY model. A more interesting possibility is to consider the two-handled double torus which we study in this paper.

The novelty of the double torus is that, as a consequence of Gauss-Bonnet theorem, it admits a metric with constant negative curvature,
\begin{equation}
\int \sqrt{g} R = 2- 2 g < 0\,,
\end{equation}
since $g = 2$. The double torus can be obtained as a quotient of the two dimensional hyperbolic plane $\mathbb{H}_2$ by a discrete subgroup of $SO(2,1)$. Another significant difference between these models and their $g=0$ and $g=1$ counterparts is the existence of preferred points in the manifold, an example of which are the Weierstrass points.  To define a Weierstrass point, first consider the set of all meromorphic functions which are holomorphic away from a specified point $P$.  $P$ is a Weierstrass point if this set contains more meromorphic functions with poles at $P$ of some specific order than are guaranteed by the Riemann-Roch theorem \cite{weierstrass}.  We will work with a particularly simple double torus corresponding to a tiling of the hyperbolic plane by regular octagons in which opposite sides are identified. The Weierstrass points corresponding to this particular identification are known \cite{weierstrass} to be the center of the octagon together with the midpoints of the sides, as well as the point determined by the eight identified vertices.  We will refer to this manifold as the regular double torus. 

In this paper we will perform Monte-Carlo simulations of the double torus model on a two dimensional square lattice with periodic boundary conditions. We will study the impact of the non-standard topological and geometric properties of this model on the phase transition.  Our results can be summarized as follows:
 \begin{itemize}
  \item
  	There is a phase transition at a finite temperature $T_c$. Numerical results are consistent with a second order phase transition, but do not exclude the possibility that the phase transition is of infinite order as in the XY model.  Our numerical results also do not exclude the possibility of additional phase transitions.
  \item For temperatures slightly below $T_c$, the spins cluster around one of six special points on the regular double torus, which can be defined as fixed points of the discrete automorphism group.  This is quite unlike the XY model, in which there are no preferred points in the target space.
  \item For temperatures somewhat above $T_c$, vortices of many different topological types appear, their numbers following thermal distributions as one would predict from treating them as free, independent excitations.
 \end{itemize}
The organization of the rest of this paper is as follows.  In section~\ref{THEORY}, we explain the general framework of lattice models with hyperbolic quotients as target spaces and specify the precise model we are interested in.  In section~\ref{RESULTS} we briefly describe our numerical methods and then explain our results, focusing on the points just summarized.  We end with a discussion including future directions in section~\ref{DISCUSSION}.

\section{Theory}
\label{THEORY}
\subsection{Hyperbolic tilings and quotient spaces}
The $n$-dimensional hyperbolic space $\mathbb{H}_n$ is a maximally symmetric Euclidean manifold of constant negative curvature. It can be embedded in $\mathbb{R}^{1,n}$ in a manifestly $SO(1,n)$ symmetric manner,
\begin{equation}\label{TwoSheeted}
 1 = X_0^2 - X_1^2 - \cdots - X_n^2\,.
\end{equation}
More precisely, $\mathbb{H}_n$ is the upper sheet of the two-sheeted hyperboloid described by  Eq. \eqref{TwoSheeted}.
We consider non-linear sigma models with target space given by the quotient space $\mathbb{H}_n/\Gamma$
where $ \Gamma $ is a discrete subgroup of the orientation-preserving isometries $SO^+(1,n)$. In the `upstairs' picture we can think of the covering space $\mathbb{H}_n$ as being tiled or tessellated by cells, each of which is related to the fundamental cell by the action of a particular element of the group $\Gamma$. We only consider orientation-preserving isometries so that the resulting topological space is orientable. Moreover, we assume that $\Gamma$ is freely acting so that the resulting quotient space does not have fixed points.

The Poincar\'{e} ball model maps the hyperboloid $\mathbb{H}_n$ to the unit ball where the induced metric is given by
\begin{equation}
ds^2 = \frac{4}{(1-r^2)^2}(dr^2 + r^2 d\Omega_{n-1}^2)\,.
\end{equation}
We will mostly focus on the $n=2$ case where the isometries $SO^+(1,2)$ are realized as fractional linear transformations acting on the unit disc in $\mathbb{C}$,
\begin{equation}
	z \longrightarrow \frac{az + \bar{c}}{cz + \bar{a}}\,,
\end{equation}
with $a,b\in\mathbb{C}$ such that $|a|^2-|c|^2 =1$.
The geodesics in the Poincar\'{e} coordinates are either diameters or arcs of circles intersecting orthogonally with the boundary of the disc. These geodesics form the edges of the regular hyperbolic polygons which tile $\mathbb{H}_2$. A tiling by regular $p$-gons with $q$ polygons meeting at each vertex exists provided that $1/p + 1/q < 1/2$. We only consider tilings with $p$ even so that the sides of each polygon can be paired. The subset $\gamma \subset \Gamma$ of group elements which pair the sides of the polygon are the generators of $\Gamma$. Note that $\gamma$ is not a group and the group $\Gamma$ is obtained by multiplying the elements of $\gamma$ in all possible ways,
\begin{equation}
	\Gamma = \{ g_1 \cdots  g_n \; | \; g_i \in \gamma  \}\,.
\end{equation}
In this way we construct a compact orientable surface with constant negative curvature. Note that this procedure will not always lead to a smooth surface. For example in the case of the $\{8,8 \}$ tessellation we obtain the genus-2 hyperbolic Riemann surface, i.e.~the regular double torus, but the surface obtained from the $\{4,5\}$ tessellation must necessarily have cusps in order to be consistent with the Gauss-Bonnet theorem.

\begin{figure}[h]
\centering
\includegraphics[width=70mm]{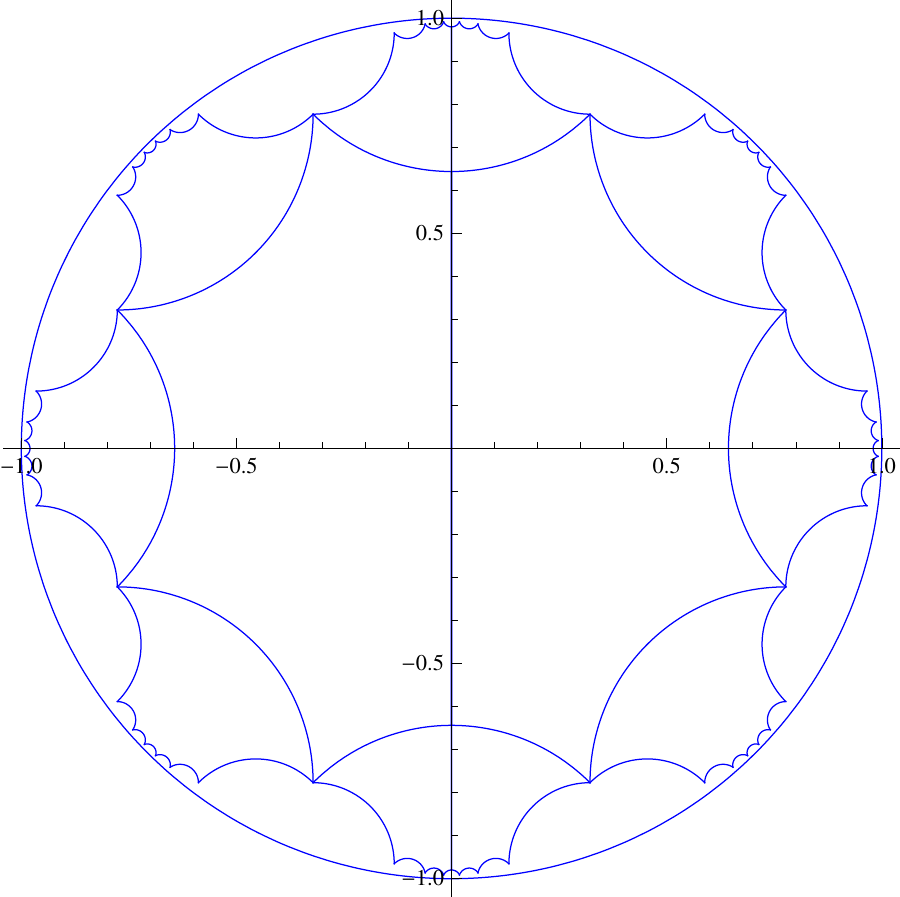}
\caption{The fundamental domain and neighboring cells for the $\{8,8\}$ tessellation.}
\label{fig:tesselation}
\end{figure}

\subsection{Lattice Hamiltonian}

The general structure of the Hamiltonians we consider is
 \begin{equation}\label{HForm}
  H = \sum_{\langle x,y \rangle} h(s_x,s_y) \,.
 \end{equation}
Here $x$ and $y$ label sites on the square lattice, and the sum is over nearest neighbors.  $s_x$ and $s_y$ are the `spins,' in other words $s_x \in M$ for each point $x$ on the lattice.  For the XY model $M=S^1$ and the standard approach is to replace it with a real variable $\theta_x$, with $\theta_x$ and $\theta_x + 2\pi$ identified.  The function $h(s_x,s_y)$ is a map $M \times M \to \mathbb{R}$ which is bounded below, usually with its lower bound attained precisely when $s_x=s_y$.  For the XY model, the standard choice is
 \begin{equation}\label{StandardH}
  h(\theta_x,\theta_y) = 1-\cos(\theta_x-\theta_y) \,.
 \end{equation}
A convenient choice for some purposes is the so-called `Villain approximation' \cite{Janke:1986ej},
 \begin{equation}\label{ProperVillain}
  h(\theta_x,\theta_y) = \min_{n \in \mathbb{Z}} (\theta_x+2\pi n-\theta_y)^2 \,,
 \end{equation}
where the sum over $n$ enforces $2\pi$ periodicity of $\theta_x$ and $\theta_y$. The natural generalization of Eq. \eqref{ProperVillain} to the quotient $\mathbb{H}_2 / \Gamma$ is
 \begin{equation}\label{ExtendedVillain}
  h(s_x,s_y) = \min_{\gamma \in \Gamma} \left[ \gamma(s_x) - s_y \right]^2 \,,
 \end{equation}
with $s_x$ and $s_y$ points on $\mathbb{H}_2 \subset \mathbb{R}^{1,2}$.  The right-hand side is non-negative because any two points on the hyperboloid $\mathbb{H}_2$ are spacelike separated.

An important feature of the Villain energy function (\ref{ProperVillain}) is that it is continuous but only piecewise smooth: there is a discontinuity in its first derivative along the locus where $\theta_x-\theta_y \equiv \pi \Mod 2\pi$.  Likewise, the generalization (\ref{ExtendedVillain}) is continuous but only piecewise smooth: For example, if $s_y$ is at the origin of the fundamental octagon, then $h(s_x,s_y)$ has discontinuities in one of its first derivatives at the boundaries of all images of that octagon.  An intrinsic coordinate system with periodically defined coordinates is not known, so it is non-trivial to give an explicit, smooth map analogous to (\ref{StandardH}).  We will therefore work strictly with the `Villain' form (\ref{ExtendedVillain}), which we may equivalently define as
 \begin{equation}\label{EquivalentH}
  h(s_x,s_y) = \min_{\gamma \in \Gamma} \left[ -2 \gamma(s_x) \cdot s_y \right] \,,
 \end{equation}
where the dot product is in the standard mostly plus flat metric on $\mathbb{R}^{1,2}$.

The equivalent forms (\ref{ExtendedVillain}) and (\ref{EquivalentH}) are not suited to computation unless we can efficiently restrict the minimization to a small subset of the elements in $\Gamma$.  For the XY model, this is easy to do: one requires $\theta_x,\theta_y \in (-\pi,\pi)$, and then the only images one needs to consider are $\theta_x$ and $\theta_x \pm 2\pi$.  We must ask: If $O \subset \mathbb{H}_2$ is the fundamental octagon, and we require $s_x,s_y \in O$, then what is the analogous subset of images $\gamma(s_x)$ that we must minimize over to be sure of finding the minimum value of $-\gamma(s_x) \cdot s_y$?

\begin{figure}[h]
\centering
\includegraphics[width=70mm]{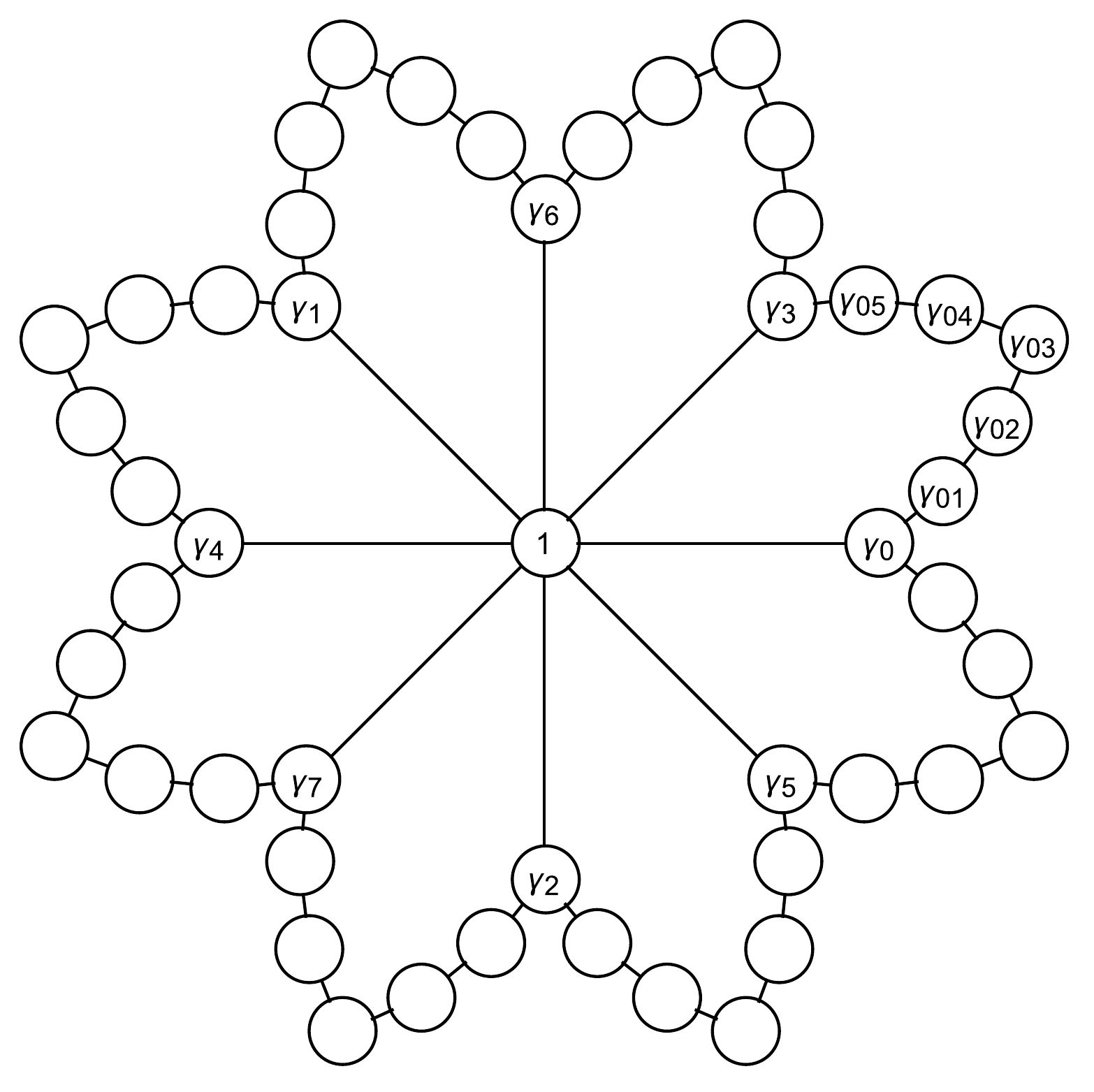}
\caption{The subgraph of the $\{8,8\}$ tiling of $\mathbb{H}_2$ corresponding to the group elements $\gamma_{jk}$.  Each node of the graph corresponds to an octagon in the tiling.  Connected nodes correspond to octagons which share an edge.  The octagons corresponding to group elements $\gamma_{01}$, $\gamma_{02}$, $\gamma_{03}$, $\gamma_{04}$, and $\gamma_{05}$ all share a vertex with the $1$, $\gamma_0$, and $\gamma_5$ octagons.}
\label{fig:graph}
\end{figure}

A sufficiently large subset of $\Gamma$ for the regular double torus is the identity element together with elements $\gamma$ such that $\gamma(O)$ touches the fundamental octagon either along a side or at a corner.   This subset, call it $\Gamma_{49}$, has $49$ such elements, which can be constructed as follows.  A standard basis 
 \begin{equation}
  \{\gamma_0,\gamma_1,\gamma_2,\gamma_3,\gamma_4,\gamma_5,\gamma_6,\gamma_7\} = \{\alpha_1,\beta_1,\alpha_2,\beta_2,\alpha_1^{-1},\beta_1^{-1},\alpha_2^{-1},\beta_2^{-1}\}
 \end{equation}
for $\Gamma$ satisfies the identity $\prod_{i=0}^7 \gamma_i = 1$.  The $49$ group elements of interest are $1$ together with
 \begin{equation}
  \gamma_{jk} \equiv \prod_{i=j}^k \gamma_{i \Mod 8} \,,
 \end{equation}
where $j < k$.  Counting the distinct $\gamma_{jk}$ is straightforward if we consider how they move us along the dual graph to the octagonal tiling of $\mathbb{H}_2$, a subgraph of which is shown in Fig.~\ref{fig:graph}.  A simple topological way to define $\Gamma_{49}$ is that it is the minimal set of generators such that an open set $S \subset \mathbb{H}_2$ can be found satisfying $\bigcup_{\gamma \in \Gamma_{49}} \gamma(O) \supset S \supset O$.

A simulation of the torus model amounts to designing a Markov chain process which results in random sampling of configurations from a probability distribution proportional to $e^{-H/T}$, where $H$ is given by (\ref{HForm}) and $h$ is given equivalently by (\ref{ExtendedVillain}) or (\ref{EquivalentH}).  To build such a process, one must be able to choose spins $s_x$ in the fundamental octagon with uniform probability with respect to the natural measure inherited from $\mathbb{H}_2$ and one must be able to evaluate all instances of $h(s_x,s_y)$, which in practice is done by restricting the minimization in (\ref{ExtendedVillain}) or (\ref{EquivalentH}) to $\gamma \in \Gamma_{49}$.

\section{Simulation results}
\label{RESULTS}

\begin{figure}[h]
\centering
\includegraphics[width=70mm]{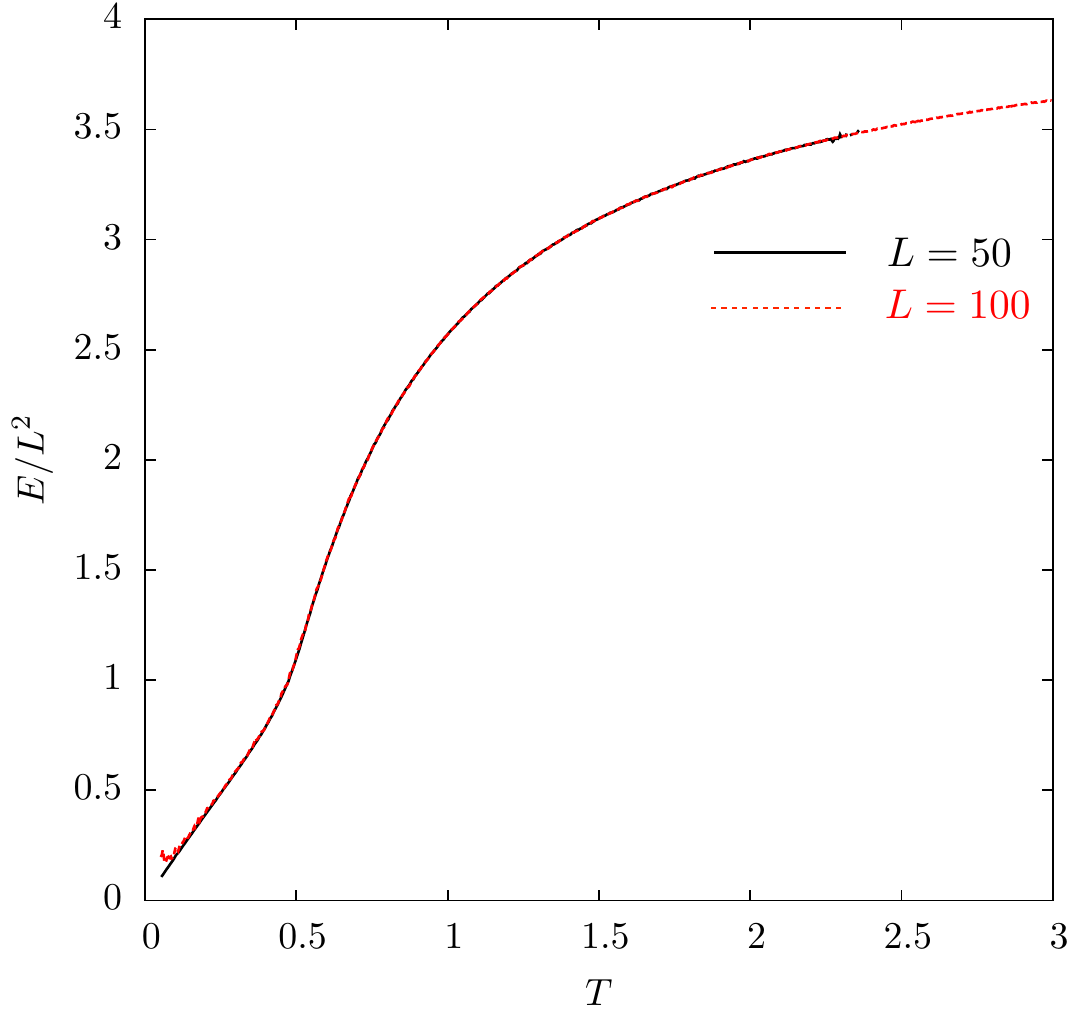}
\includegraphics[width=70mm]{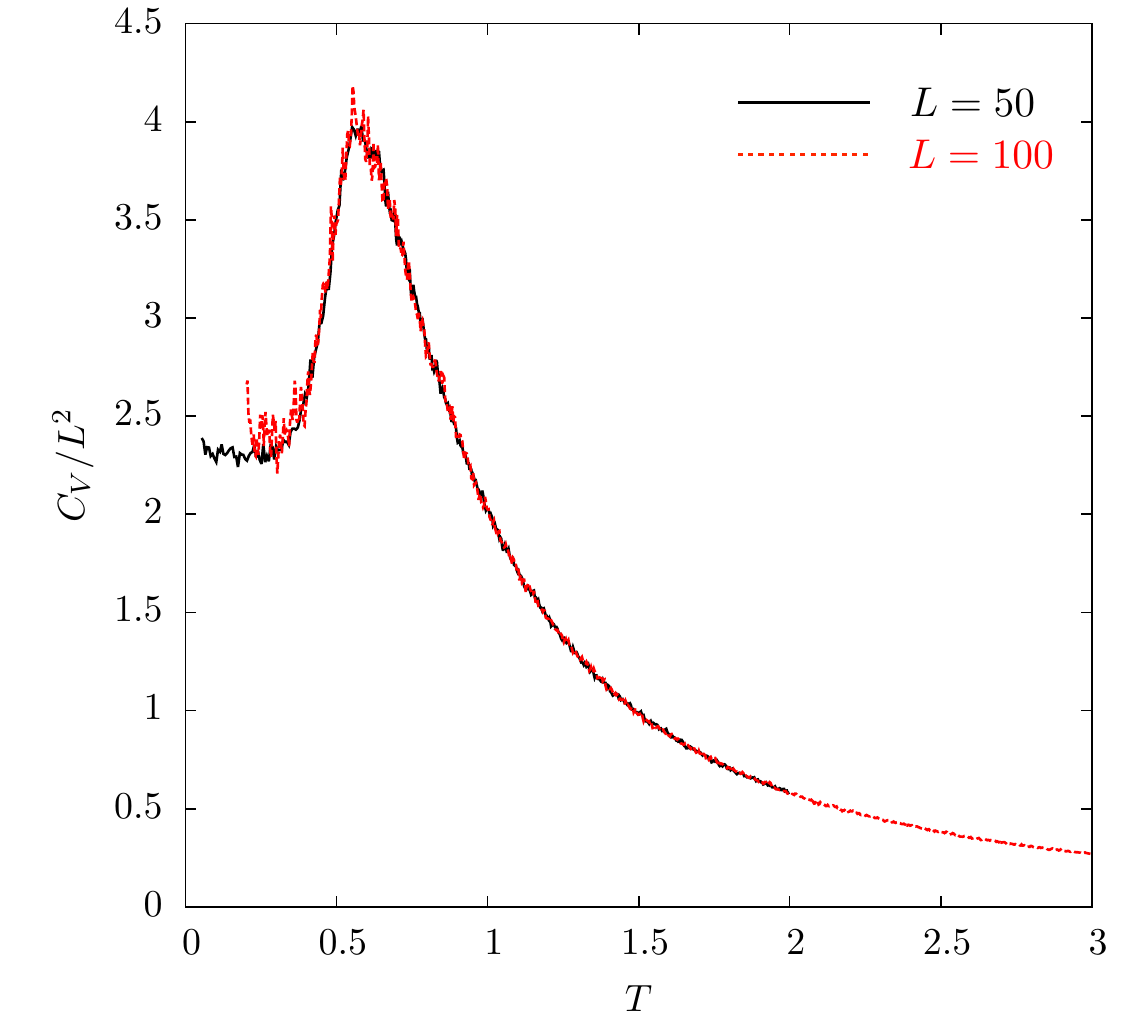}
\caption{Average energy (left) and heat capacity (right) for the $\{8,8\}$ model in the Villain approximation. There is clear evidence of a phase transition at $T\simeq 0.6$.}
\label{fig:cv}
\end{figure}

\begin{figure}[h]
\centering
\includegraphics[width=6in]{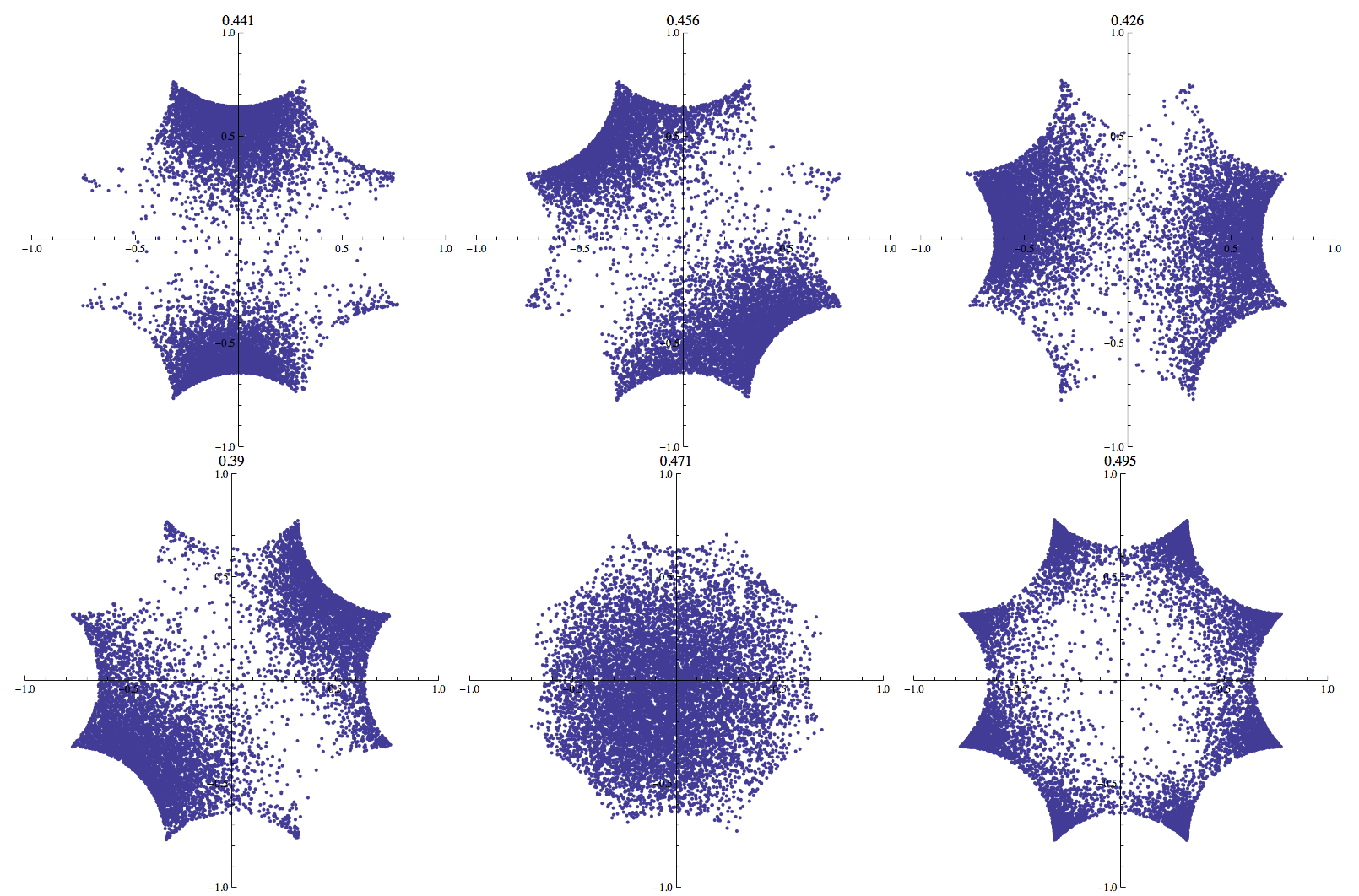}
\caption{Preferred spin configurations below critical temperature.}\label{fig:clustering}
\label{preferredspins}
\end{figure}

For temperatures $T$ slightly smaller than the critical temperature where the specific heat is maximized (Fig.~\ref{fig:cv}), we notice a surprising clustering of spins around one of the six Weierstrass points (Fig.~\ref{fig:clustering}).\footnote{We observe only one pronounced local maximum for the specific heat.  Thus we only have evidence for one critical temperature where a phase transition may occur.  It is interesting, however, that there are many types of vortices, and it is not impossible that there could be many phase transitions as a result.}  Infrared fluctuations must eventually cause the system to explore all possible regions of phase space, so for sufficiently large lattices we would expect to see a distribution of spins which is democratic among the Weierstrass points at any fixed temperature.  Thus the interesting point is that a spatial correlation length is large enough near the critical temperature so that essentially our whole lattice clusters in the vicinity of one Weierstrass point.

\begin{figure}[h]
\centering
\includegraphics[width=70mm]{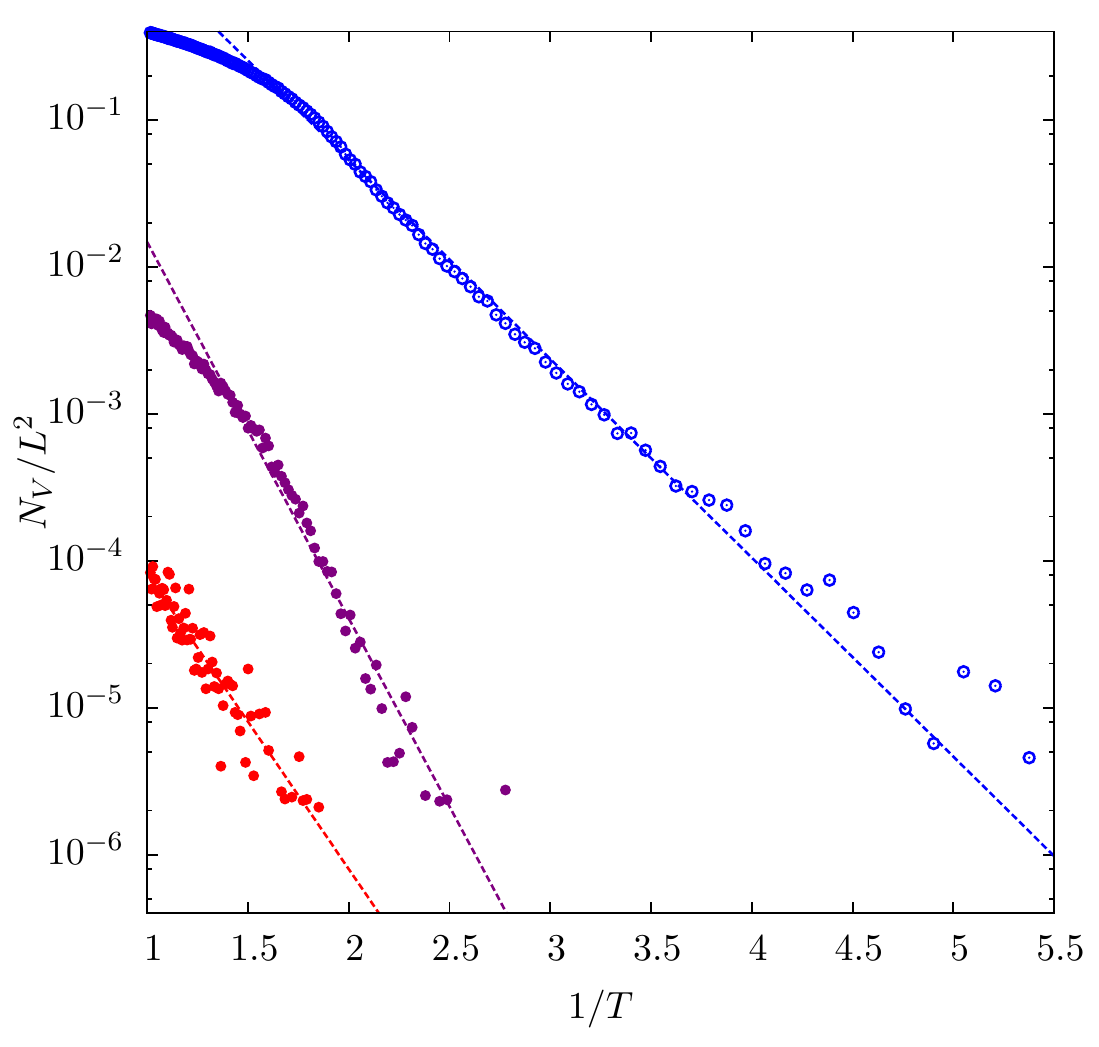}
\includegraphics[width=70mm]{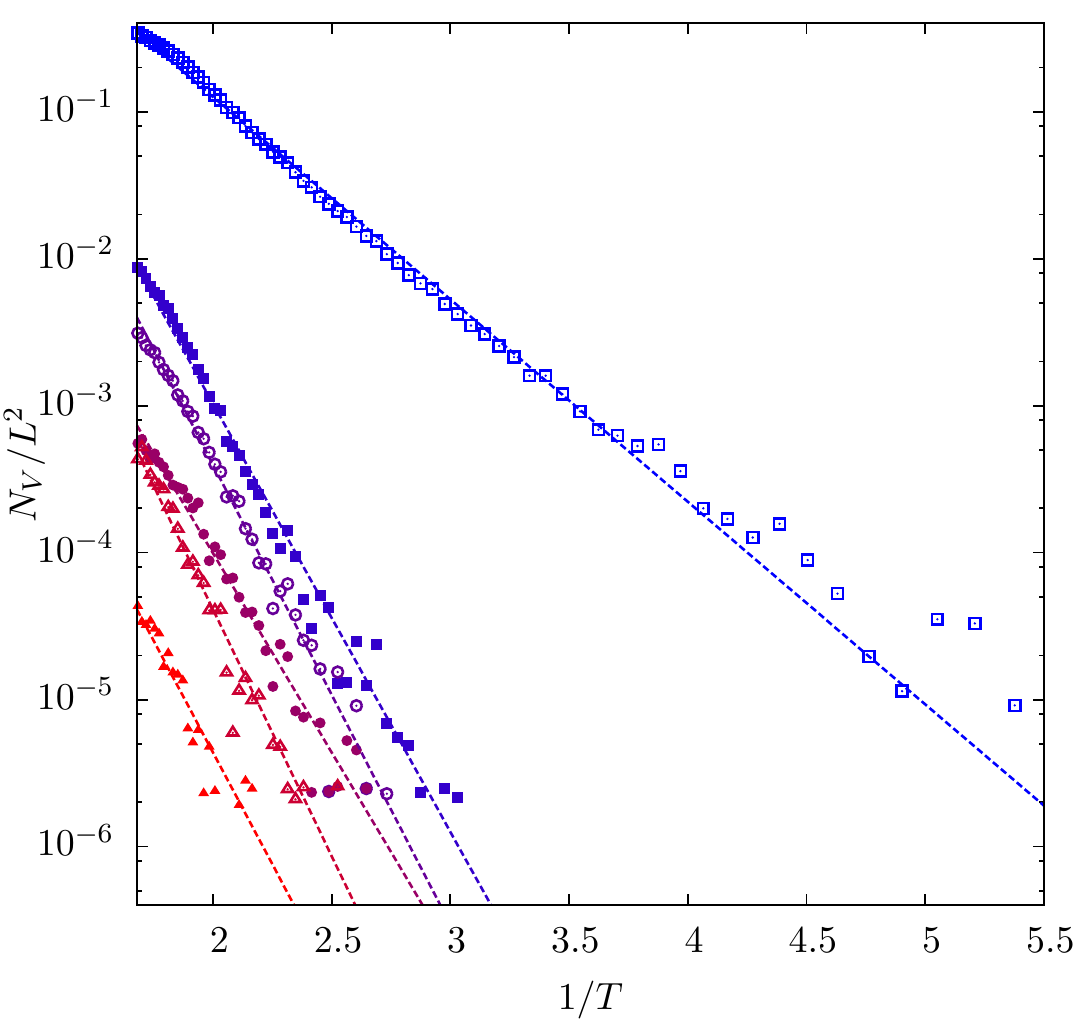}
\includegraphics[width=70mm]{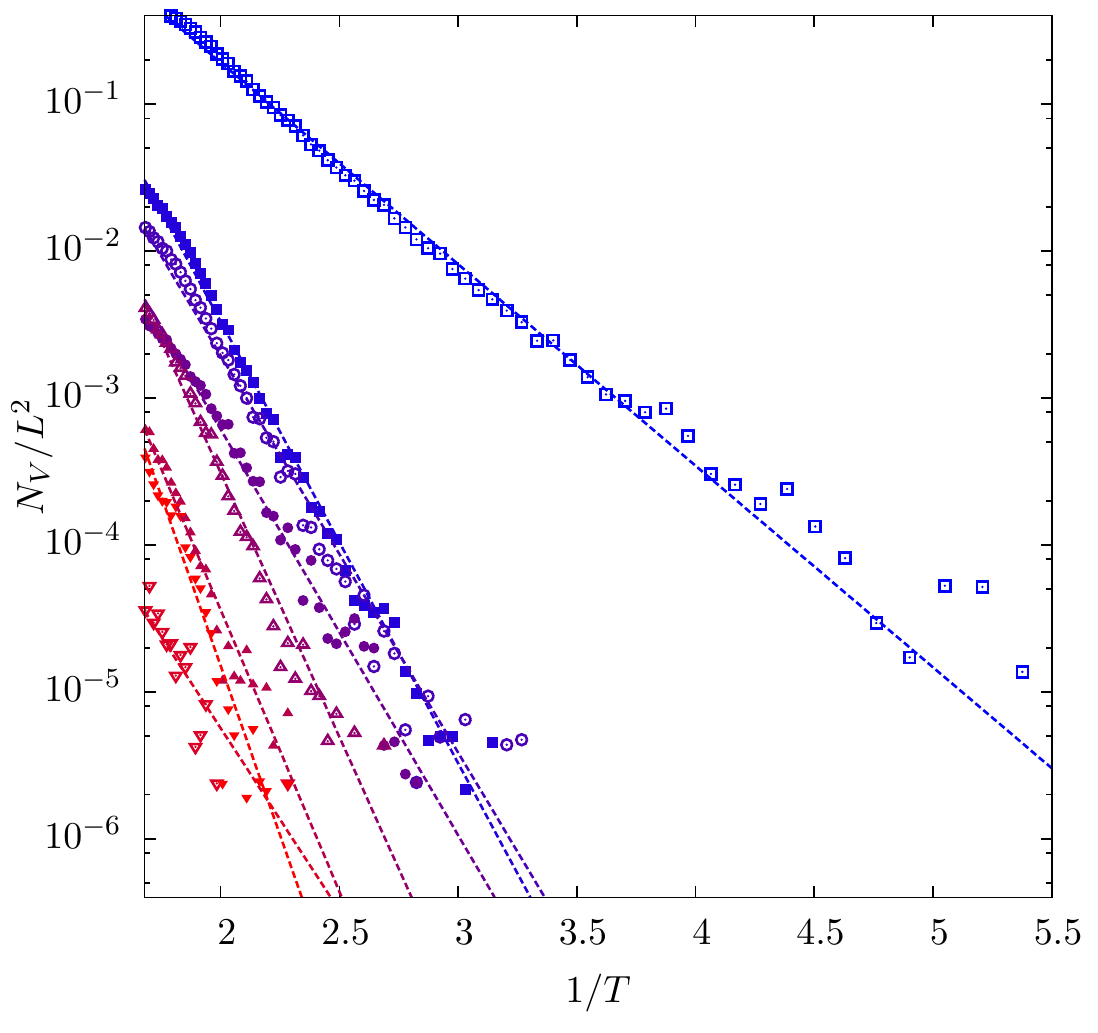}
\caption{Number vortices of fixed trace using plaquettes of size $1\times 1$ (left), $2\times 2$ (right) and $3\times 3$ (center).}\label{fig:Nv}
\label{NvSVs}
\end{figure}

\begin{table}[h]
\begin{center}
\begin{minipage}{.5\textwidth}
\ctable[
    %caption = {$S=1$},
    %label = tab:table1,
    pos= H,
]{ccc}{} {
\multicolumn{3}{ c }{$1\times1$ plaquettes}\\
\FL
$N_0$ & $E$ & Trace
\ML
 25.0 & 3.09 & 22.3 \NN
 7.51 & 6.05 & $135.$ \NN
 $4.18 \cdot 10^{-3}$ & 4.18 & $341.$ \LL
}
\ctable[
    %caption = {$S=2$},
    %label = tab:table2,
    pos= H,
]{ccc}{} {
\multicolumn{3}{ c }{$2\times2$ plaquettes}\\
\FL
$N_0$ & $E$ & Trace 
\ML
 58.5 & 3.11 & 22.3 \NN
 558. & 6.62 & 135. \NN
 249. & 6.74 & 341. \NN
 13.0 & 5.93 & 453. \NN
 100. & 7.36 & 791. \NN
 12.4 & 7.41 & $1.22 \cdot 10^3$ \NN
 \light{4.54} &  \light{7.16} & \light{$ 1.43 \cdot 10^3$} \NN
 \light{0.0233} & \light{5.15} & \light{$1.99 \cdot 10^3$} \NN
 \light{0.755} & \light{6.56} &  \light{$2.65 \cdot 10^3$} \NN
 \light{0.0749} & \light{5.86} & \light{$3.40 \cdot 10^3$} \LL
}

\end{minipage}%THIS COMMENT IS SIGNIFICANT why?
\begin{minipage}{.5\textwidth}
\ctable[
    %caption = {$S=3$},
    %label = tab:table3,
    pos= H,
]{ccc}{} {
\multicolumn{3}{ c }{$3\times3$ plaquettes}\\
\FL
$N_0$ & $E$ & Trace
\ML
 78.9 & 3.06 & 22.3\NN
 $3.41 \cdot 10^3 $ & 6.92 & 135.\NN
 481. & 6.21 & 341.\NN
 185. & 6.32 & 453.\NN
 $2.16 \cdot 10^3 $ & 7.91 & 791.\NN
 396. & 8.09 & $1.22 \cdot 10^3 $\NN
 $1.06 \cdot 10^3 $ & 9.00 & $ 1.43 \cdot 10^3$\NN
 1.76 & 6.48 & $ 1.99 \cdot 10^3$ \NN
 \light{155.} & \light{7.98} & \light{$2.65 \cdot 10^3$}\NN
 \light{31.1} & \light{7.86} & \light{$3.40 \cdot 10^3$}\NN
 \light{249.} & \light{8.49} & \light{$3.73 \cdot 10^3$}\NN
 \light{1.63} & \light{6.29} & \light{$4.62 \cdot 10^3$}\NN
 \light{20.8} & \light{8.08} & \light{$5.59 \cdot 10^3$}\NN
 \light{2.53} & \light{6.95} & \light{$6.02 \cdot 10^3$}\NN
 \light{29.3} & \light{7.65} & \light{$7.13 \cdot 10^3$}\LL
}
\end{minipage}
\end{center}
\caption{Fit parameters for plaquettes of size $1\times1$, $2\times2$ and $3\times3$. The emphasized values correspond to the data in Fig.~\ref{fig:Nv}.}
\label{table:alltables}
\end{table}

Above the critical temperature a topological phase transition occurs as a result of of vortex proliferation. Unlike the XY-model where the winding of a loop of spins is characterized by an integer, the winding of spins in the double torus model is characterized by an operator, which can be determined in the following way: For any consecutive pair of spins $s_1$ and $s_2$ in the loop we find the image of $s_2$ that is closest to $s_1$. This amounts to finding the operator $O_{s,t}\in\Gamma_{49}$ that minimizes the distance $s \cdot (Ot)$. If the loop consists of spins $s_1, \ldots, s_n$, the operator corresponding to the loop is
\begin{equation}
	O_{\rm loop} = O_{s_1,s_2} \cdots O_{s_{n-1},s_n}  O_{s_n,s_1} \,.
\end{equation} 
We map this operator to a number by taking the matrix trace.
\

The fact that the winding is characterized by a non-abelian matrix rather than a number suggests that we should consider loops which are larger than the elementary $1\times 1$ plaquette. If we define vortices with respect to a single plaquette, then only finitely many possible group elements can appear in $O_{\rm loop}$ but we may see a bigger set of group elements if we consider larger loops. Indeed, this intuition is corroborated by simulations (see Fig.~\ref{fig:Nv}) in which we see that the number of vortices of fixed trace increases as we increase the loop size from $1\times 1$ to $3\times 3$. The logarithmic scale on these plots indicates that vortices are proliferating approximately according to Boltzmann statistics; that is,
\begin{equation}
	N_V = N_0 \exp(-E/T)  \label{BoltzLaw}
\end{equation} 
for some $N_0$ and $E$. Moreover, the vortices appear for all temperatures, not just for $T>T_c$.

\begin{figure}[h]
\centering
\includegraphics[width=70mm]{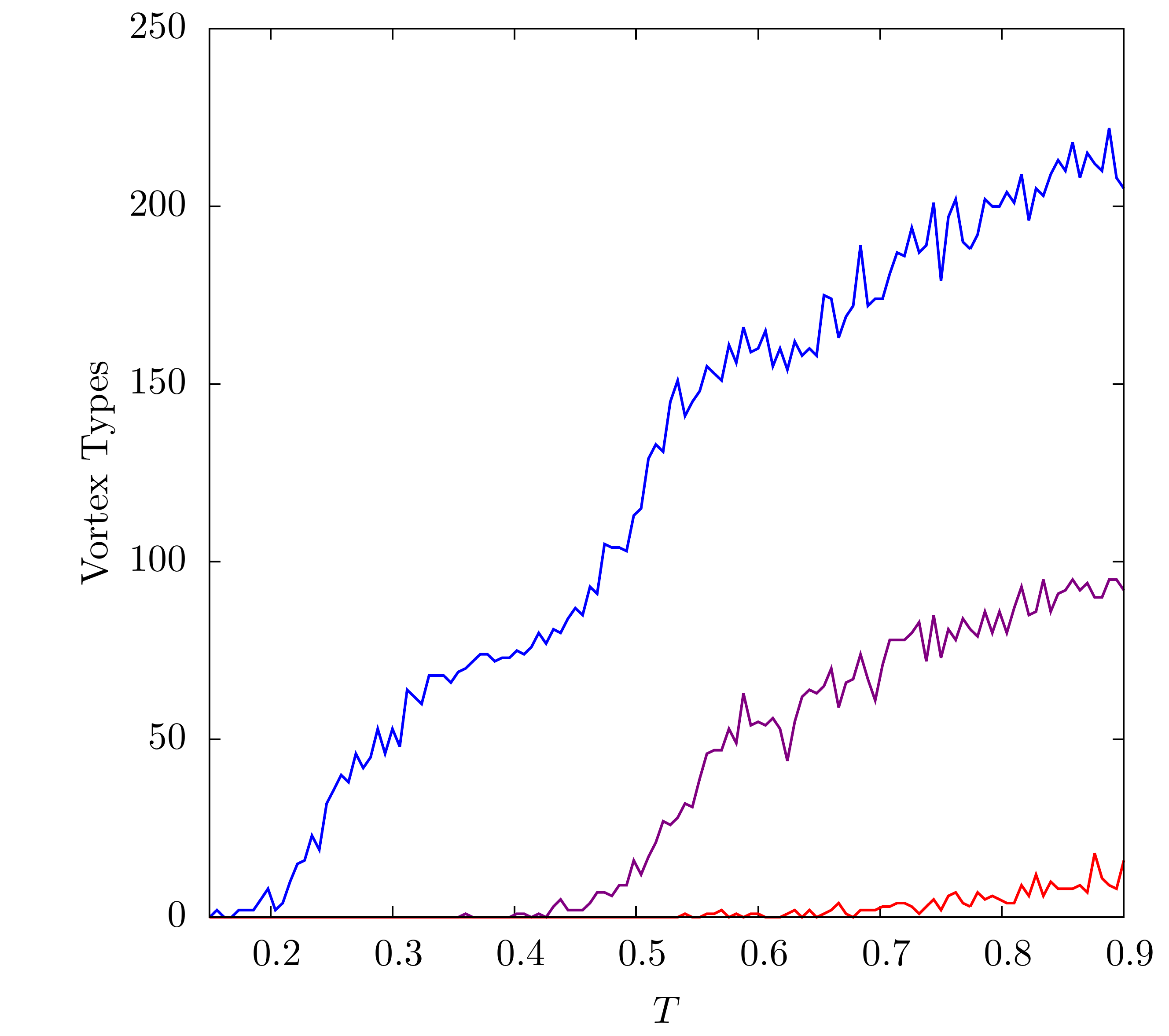}
\caption{The number of different $O_{\text{loop}}$ elements for fixed trace observed in the system as a function of temperature from trace 22.37 (blue) to trace 340 (red).}
\label{vortypes}
\end{figure}

A closer examination of the occurrence of vortices is worthwhile.  We refer to Fig.~\ref{fig:Nv} and Table~\ref{table:alltables}.  The main features to note are:
 \begin{itemize}
  \item The exponential law (\ref{BoltzLaw}) persists up to temperatures comparable to $T_c$.  At larger temperatures, we see some evidence of saturation, where the number of vortices per site becomes greater than $1/10$, and continuing to follow (\ref{BoltzLaw}) would eventually conflict with the limit $N_V / L^2 < 1$.  Provisionally then, we think of (\ref{BoltzLaw}) as a dilute gas approximation.
  \item We found it useful to distinguish vortices based on the trace of their monodromy matrix $O_{\rm loop}$.  Bigger $\tr O_{\rm loop}$ presumably means a larger vortex with bigger energy.  But we cannot conclude that all vortices with the same $\tr O_{\rm loop}$ have the same energy.  Most likely, each trace class includes vortices of different energies, and the vortices with the lowest energies dominate $N_V$ in that particular trace class. To see this explicitly we plot in Fig.~\ref{vortypes} the number of different $O_{\rm loop}$ observed with constant $\tr O_{\rm loop}$ as a function of temperature for $1\times 1$ plaquettes. We see that as $\tr O_{\rm loop} = 22.37$ begins to proliferate the number of observed $O_{\rm loop}$ increases in turn as expected. However, we also see that as $\tr O_{\rm loop} = 134.9$ vortices begin to appear a second, unexpected, jump in the number of $\tr O_{\rm loop}=22.37$ vortices also occurs. We interpret this increase to hint that a population of $\tr O_{\rm loop}=22.37$ have higher energy. Thus $\tr O_{\rm loop}$ is correlated with energy but does not uniquely determine it.
  \item The energies $E$ determined by fitting $N_V$ in a given trace class to the Boltzmann form (\ref{BoltzLaw}) do not change much when we go from $1 \times 1$ plaquettes to $2 \times 2$ to $3 \times 3$.  Typical changes were on order $5\%$ to $10\%$ or less.  The energy for $\tr O_{\rm loop} = 340.7$ changes more than this between $1 \times 1$ and $2 \times 2$, but in this case the sample size for $1 \times 1$ plaquettes is very small.  The robustness of $E$ encourages the view that we can treat vortices in a dilute gas approximation.  But we should be a little cautious about values of $E$ since the fits are sometimes shaky due to noisy data and a limited range of $1/T$.
  \item The prefactor $N_0$, which we think of as related to the exponential of the fugacity, is conspicuously different for some trace classes from $1 \times 1$ plaquettes to $2 \times 2$ to $3 \times 3$.  When $N_0$ rises drastically with plaquette size, as it does for trace $340.7$ vortices from $1 \times 1$ to $2 \times 2$ plaquettes, we presume that a contributing effect is that the characteristic size of the vortex is too big for the smaller plaquette size to have a significant chance of capturing the full vorticity.  Monotonic increase in $N_0$ is to be expected as plaquette size increases, because if the vortex is small, a large plaquette will enclose it completely at several different positions, all of which contribute to $N_0$.
\end{itemize}

We suspect that a low-temperature expansion in terms of a dilute gas of vortices of many types can be used to account for much of the dynamics up to $T_c$.  It would be interesting to pursue this further because it could be a low-dimensional analog to the hadron gas treatment of the low-temperature phase of QCD, inspired originally by Hagedorn's statistical bootstrap approach \cite{Hagedorn:1965st}.  Optimistically, one might try to estimate $T_c$ in terms of the growth in the number of different vortices with energy.  The number of different vortex types increases exponentially with the length on $\mathbb{H}_2$ between an initial point $s$ and its image $O_{\rm loop} s$.  (This is essentially the statement that the area enclosed by a circle in $\mathbb{H}_2$ increases exponentially with its radius.)  So it is not implausible that the number of different vortex types also increases exponentially with energy, facilitating a Hagedorn-style argument where a ``vortex gas'' eventually reaches a maximum possible temperature, which is $T_c$.  However, with our present numerical results, we cannot go very far in developing such a vortex gas model, for two main reasons.  First, we don't have a clear notion of the energy of a vortex; certainly the trace $\tr O_{\rm loop}$ is only very loosely correlated with the energy $E$ as obtained from a fit of $N_V$ to the Boltzmann form (\ref{BoltzLaw}) for vortices in a given trace class.  Moreover, the standard Kosterlitz-Thouless treatment of the XY model discourages us from thinking that the energy of a single vortex is well-defined in isolation.  Second, the coefficients $N_0$ are not really accessible from numerics.  Ideally they should correlate with the number of vortices of a given energy.

A simple alternative point of view is that $T_c$ occurs naturally when the dilute gas approximation for the smallest vortices breaks down, and the larger vortices have at most a modest effect on the thermodynamics.  Referring to the $1 \times 1$ plaquette results in Fig.~\ref{fig:Nv}, the dilute gas approximation for the smallest vortices does break down for $T$ only modestly below $T_c$.

\section{Discussion and future directions}
\label{DISCUSSION}
We have considered one of an infinite number of tilings of the hyperbolic plane $\mathbb{H}_2$. The $\{8,8\}$ is simplest tiling which yields a smooth target manifold but it is straightforward to generalize to hyperbolic Riemann surfaces of higher genera. It will also be interesting to consider quotients of hyperbolic space $\mathbb{H}_n$ with $n>2$. In addition, it would be interesting to consider deformations in the moduli space of the double torus which correspond to changing the geodesic lengths of the various cycles. Under such conditions one can expect a splitting of the phase transition into different critical temperatures corresponding to each cycle.

A natural question to ask is whether the sigma models studied in this paper have well-defined UV completions. A general nonlinear sigma model is a theory of maps $\mathbb{R}^2 \longrightarrow T$ with Euclidean action functional given by
\begin{equation}
		S = \frac{1}{2\alpha_0}\int d^2 x \, g^{\mu\nu} \partial_\alpha X_\mu \partial^\alpha X_\nu \, .
	\end{equation}
The beta function for this theory can be computed perturbatively in the coupling parameter $\alpha_0$. Assuming that $T$ is an $n$-dimensional maximally symmetric space for simplicity,
	\begin{align}
		\beta_{\mu\nu}
			& =  \alpha'R_{\mu\nu}  + \frac{\alpha'^2}{2} R_{\mu\lambda\sigma\rho}R_{\nu}^{\phantom{\nu}\lambda\sigma\rho} + \mathcal{O}(\alpha'^3), \quad \quad \alpha' \equiv \frac{\alpha_0}{2\pi}, \\
			\beta(\alpha')
				& = \frac{\partial \alpha'}{\partial \log \mu} = -\frac{\alpha'}{n}g^{\mu\nu}\beta_{\mu\nu}.
\end{align}
Here $\mu$ and $\nu$ are curved indices for the target manifold $T$ and $\alpha$ is a flat index for the worldsheet $\mathbb{R}^2$. Taking $T = \mathbb{H}_{n}$, the one-loop beta function is given by $\beta(\alpha') = (n-1)\alpha'^2$, which naively suggests a Landau pole in the UV. It is conceivable, however, that the higher derivative terms can balance the one-loop contribution and drive the theory to a conformal fixed point \cite{Friess:2005be,Michalogiorgakis:2006jca}. There also exist arguments \cite{Polyakov:2007mm} that models defined on compact targets with negative curvature $\mathbb{H}_n/\Gamma$ should possess conformal fixed points which arise from the competition between infrared freedom at weak coupling and the discrete spectrum of the Laplacian at strong coupling. It is therefore an important open problem to verify if the transition seen in simulations is of the second-order type, and to ascertain if critical fluctuations are described by a conformal field theory. Answering this delicate question may require moving beyond the Villain approximation, by finding an exact embedding of hyperbolic double torus.

In this paper we have focused on the models with constant negative curvature and compact target $\mathbb{H}_n/\Gamma$ but it is also interesting to consider models on the non-compact space $\mathbb{H}_n$. The non-compactness of the target makes it difficult to define correlation functions and a non-standard basis of observables may be required \cite{Polyakov:2014rfa}.

Finally, the modern conformal bootstrap techniques have so far only been applied to theories with compact symmetry groups. It would be very interesting to extend them to the non-compact groups.

\section{Acknowledgements}
J.S., Z.H.S. and S.S.S. would like thank Randall Kamien, Hernan Piragua  and Alexander Polyakov for discussions. B.S. would like to thank Hirosi Ooguri for useful discussions, and the Institute for Advanced Study, Princeton University, and the Simons Center for Geometry and Physics for hospitality. B.S. also gratefully acknowledges support from the Simons Summer Workshop 2015, at which part of the research for this paper was performed. J.S. is supported in part by NASA ATP grant NNX14AH53G. Z.H.S. is supported in part by DOE Grant DOE-EY-76-02-3071. S.S.S. is supported by DOE DE-FG02-05ER46199. B.S. is supported in part by the Walter Burke Institute for Theoretical Physics at Caltech and by U.S. DOE grant DE-SC0011632.

\appendix
\section{Coordinate systems}
The following coordinate system covers the upper sheet of $\mathbb{H}_2$ in $\mathbb{R}^{1,2}$,
\begin{align}
	X^0
		& = \cosh \rho\,, \\
	X^1
		& = \sinh \rho \cos\phi\,, \\
	X^2
		& = \sinh \rho \sin\phi\,.
\end{align}
The induced metric and area element are given by
\begin{equation}
	ds^2 = d\rho^2 + \sinh^2\rho d\phi^2\,, \quad dA = \sinh \rho d\rho d\phi\,.
\end{equation}
The form of the area element implies that we should choose $\phi$ uniformly in $[0,2\pi]$ and choose the cumulative distribution function $v$ to be uniformly distributed in $[0,1]$, where
\begin{equation}
	v = \frac{\int_0^\rho \sinh\rho' d\rho'}{\int_0^{\rho_c} \sinh\rho' d\rho' } = \frac{\cosh \rho - 1}{\cosh\rho_c -1}\,.
\end{equation}
Therefore
\begin{equation}
	\rho = \cosh^{-1}(1-v+v\cosh\rho_c)\,.
\end{equation}
Defining $x = \sinh\rho\cos\phi$ and $y = \sinh\rho\sin\phi$ we obtain the alternative parametrization
\begin{align}
	X^0
		& = \sqrt{1+x^2+y^2}\,, \\
	X^1
		& = x\,, \\
	X^2
		& = y\,.
\end{align}
The Poincar\'{e} disc model is obtained by defining $r = \tanh(\rho/2)$. Then
\begin{equation}
	ds^2 = \frac{4}{(1-r^2)^2}(dr^2 + r^2d\theta^2)\,.
\end{equation}
and the relationship with the rectangular coordinates is
\begin{align}
	r
		& = \tanh\left[\frac{1}{2}\cosh^{-1}(\sqrt{1+x^2+y^2})\right]\,, \\
	\theta
		& = \tan^{-1}\left(\frac{y}{x}\right)\,.
\end{align}

\section{Parametrization of hyperbolic polygons}
The solutions to the geodesic equation on the Poincar\'{e} disc are given by \cite{Nazarenko}
\begin{eqnarray}\label{xy}
x(s)&=&\frac{\cos{\phi}\cosh{s}-R\sin{\phi}\sinh{s}}{\sqrt{1+R^2}\cosh{s}+R}\,,
\nonumber\\
y(s)&=&\frac{R\cos{\phi}\sinh{s}+\sin{\phi}\cosh{s}}{\sqrt{1+R^2}\cosh{s}+R}\,,
\end{eqnarray}
where $s\in(-\infty,+\infty)$. These functions define  arcs on the Poincar\'{e} disc with the radius $R$ and the centers at the point $x_0=\sqrt{1+R^2}\cos{\phi}$, $y_0=\sqrt{1+R^2}\sin{\phi}$, lying beyond the unit disc. In order to draw a the fundamental polygon corresponding to the $\{p,q\}$ tessellation, we need to fix the angle $\alpha$ between any two neighboring lines from the center of the disk to the vertices to be $\alpha=2\pi/p$ and the distance $a$ of all the vertices from the center of the disk to be such that the interior angles are $2\pi/q$. We will assume the corners of the polygon are at the points $a\exp{(ik\pi/2)}$, $a\exp{i(\alpha+k\pi/2)}$. The lines connecting these points are given by the solutions to the geodesic solutions specified above. The $p$ geodesics corresponding to the $p$ exterior circles are completely specified by radii $R_\pm$ and angles $\phi_\pm+k\pi/2$ , given by \cite{Nazarenko}
\begin{equation}\label{R-phi}
R_\pm=\frac{1}{2a}\sqrt{T_\pm^2+(1-a^2)^2}\,,\qquad
\phi_\pm=\arctan{\left[\left(\frac{T_\pm}{1+a^2}\right)^{\pm1}\right]}\,, \qquad T_\pm = a^2 \pm \tan(\alpha - \pi/4)\,.
\end{equation}
For the $\{8,8 \}$ tessellation we require $a = 2^{-1/4}$ and $\phi_\pm$ with $k =0,1,3$.

\end{document}